# Exact solution of Schroedinger equation
# in the case of reduction to Riccati type of ODE.

**Sergey V. Ershkov**

Institute for Time Nature Explorations,

M.V. Lomonosov's Moscow State University,

Leninskie gory, 1-12, Moscow 119991, Russia

e-mail: sergej-ershkov@yandex.ru

**Keywords**: Schroedinger equation, Riccati ODE, paraxial approximation.

A new type of solution for the full 3+1 dimensional space-time Schroedinger equation is presented here. We consider elegant presentation of the exact solution in a spherical coordinate system, along with the assuming of separation of the two angular co-ordinates from the radial and time variables. The separation of variables follows from an assumed product form of the full potential function, which should allow us to reduce the Schroedinger Eq. to Riccati ODE in stationary case.

If the angular dependence is constant, then the azimuthal dependence is linear and the polar dependence is logarithmic. With this reduction the remaining partial differential equation links the radial and time dependence which is, in effect, the standard time-dependent spherical radial Schroedinger equation. Besides, we obtain that the time-depended 2-angles solutions exist if the axis of preferential direction of wave propagation is similar to the time arrow in mechanical processes. The other possibility is the appropriate phase-shifting of the time-depended solutions.

A *paraxial* type approximation is obtained for the solution in the limit in which the polar angle approaches zero.

# 1. Introduction.

The full 3+1 dimensional space-time, non-relativistic Schroedinger equation for the particle of mass *m*, evolving over time *t* in the presence of a potential *V*, should be presented in a spherical coordinate system $R, \theta, \varphi$ as below [1-3] (under the proper initial conditions):

$$-\frac{\hbar^2}{2m} \Delta \psi + U \psi = i\hbar \frac{\partial \psi}{\partial t}, \qquad (1.1)$$

- here $\psi$ - is the wavefunction of particle in a position space, $\psi = \psi(R, \theta, \varphi, t)$; function $U = (V - E)$, where *V* - is the potential of the particle in position space [4], *E* - is the total energy of quantum system, $U = U(R, \theta, \varphi, t)$; $i = \sqrt{-1}$ - is the imaginary unit, $\hbar$ – is the reduced Planck constant.

Besides, in a spherical coordinate system [5]:

$$\Delta \psi = \frac{\partial^2 \psi}{\partial R^2} + \frac{2}{R}\frac{\partial \psi}{\partial R} + \frac{1}{R^2 \sin^2 \theta}\frac{\partial^2 \psi}{\partial \varphi^2} + \frac{1}{R^2}\frac{\partial^2 \psi}{\partial \theta^2} + \frac{1}{R^2} ctg\,\theta \frac{\partial \psi}{\partial \theta}.$$

Let us search for solutions of equation (1.1) in a form below:

$$\psi(R, \theta, \varphi, t) = \psi_1(R, t) \cdot \psi_2(\theta, \varphi), \qquad (1.2)$$
$$U(R, \theta, \varphi, t) = U_1(R, t) \cdot U_2(\theta, \varphi),$$

- where $\theta \neq 0$. Then having substituted expressions (1.2) into Eq. (1.1), we should obtain

$$\frac{1}{\psi_1(R,t)}\frac{\partial^2\psi_1(R,t)}{\partial R^2} + \frac{2}{R}\frac{1}{\psi_1(R,t)}\frac{\partial\psi_1(R,t)}{\partial R} + \frac{1}{R^2\sin^2\theta}\frac{1}{\psi_2(\theta,\varphi)}\frac{\partial^2\psi_2(\theta,\varphi)}{\partial\varphi^2} +$$

$$+ \frac{1}{R^2}\frac{1}{\psi_2(\theta,\varphi)}\frac{\partial^2\psi_2(\theta,\varphi)}{\partial\theta^2} + \frac{ctg\,\theta}{R^2}\frac{1}{\psi_2(\theta,\varphi)}\frac{\partial\psi_2(\theta,\varphi)}{\partial\theta} + \frac{2m}{\hbar}\frac{i}{\psi_1(R,t)}\frac{\partial\psi_1(R,t)}{\partial t} = \frac{2m}{\hbar^2}U_1(R,t)\cdot U_2(\theta,\varphi)\,,$$

- or

$$\frac{1}{\psi_1(R,t)}\frac{\partial^2\psi_1(R,t)}{\partial R^2} + \frac{2}{R}\cdot\frac{1}{\psi_1(R,t)}\frac{\partial\psi_1(R,t)}{\partial R} + \frac{2m}{\hbar}\frac{i}{\psi_1(R,t)}\frac{\partial\psi_1(R,t)}{\partial t} - \frac{2m}{\hbar^2}U_1(R,t)\cdot U_2(\theta,\varphi) =$$

(1.3)

$$= -\frac{1}{R^2}\cdot\left\{\frac{1}{\sin^2\theta}\frac{1}{\psi_2(\theta,\varphi)}\frac{\partial^2\psi_2(\theta,\varphi)}{\partial\varphi^2} + \frac{1}{\psi_2(\theta,\varphi)}\frac{\partial^2\psi_2(\theta,\varphi)}{\partial\theta^2} + \frac{ctg\,\theta}{\psi_2(\theta,\varphi)}\frac{\partial\psi_2(\theta,\varphi)}{\partial\theta}\right\}$$

We should also note that if the source of potential of the particle coincides with the origin of spherical coordinate system (*besides, such a source generates the field of central symmetry structure*), then the condition $U_2(\theta,\varphi)$ = const ≡ 1 should be chosen for Eq. (1.3); we will consider only such a case here and below.

## 2. Exact solution.

According to the method used for the obtaining of exact solutions of the Helmholtz equation [6], let us choose in Eq. (1.3):

$$\frac{1}{\sin^2\theta}\frac{1}{\psi_2(\theta,\varphi)}\frac{\partial^2\psi_2(\theta,\varphi)}{\partial\varphi^2} + \frac{1}{\psi_2(\theta,\varphi)}\left(\frac{\partial^2\psi_2(\theta,\varphi)}{\partial\theta^2} + ctg\,\theta\frac{\partial\psi_2(\theta,\varphi)}{\partial\theta}\right) = C = const \quad (2.1)$$

Besides, we should note especially the case below ($C = 0$; $\theta \in (0, \pi)$):

$$\frac{\partial^2 \psi_2(\theta,\varphi)}{\partial \theta^2} + ctg\,\theta \cdot \frac{\partial \psi_2(\theta,\varphi)}{\partial \theta} = 0 \quad \Rightarrow \quad \frac{\partial \psi_2(\theta,\varphi)}{\partial \theta} \cdot \sin\theta = f_1(\varphi)$$

$$\Rightarrow \quad \psi_2(\theta,\varphi) = f_1(\varphi) \cdot \ln\left(tg\frac{\theta}{2}\right) \quad \Rightarrow \quad f_1(\varphi) = C_1 \varphi + C_2 \qquad (2.2)$$

- here $C_1$, $C_2$ – are some constants (according to the initial conditions), $C_1 \neq 0$.

If $C = 0$, the equality (1.3) under assumption (2.1) could be reduced as below:

$$\frac{\partial^2 \psi_1(R,t)}{\partial R^2} + \frac{2}{R} \cdot \frac{\partial \psi_1(R,t)}{\partial R} - \left(\frac{2m}{\hbar^2} \cdot U_1(R,t)\right) \cdot \psi_1(R,t) = -\frac{2m}{\hbar} i \cdot \frac{\partial \psi_1(R,t)}{\partial t} \qquad (2.3)$$

- which is the standard time-dependent spherical radial Schroedinger equation.

In stationary case ($\partial/\partial t = 0$), the equation (2.3) above is known to be of *Riccati* type [5] in regard to variable $R$. In the non-stationary case $\partial/\partial t \neq 0$, such an equation could be solved analytically only if:

1) $\partial/\partial t \sim \partial/\partial R$ - it means that the **R** axis represents a preferential direction similar to the time arrow in mechanical processes [7];

2) $\partial \psi_1(R,t)/\partial t \sim \psi_1(R,t) \quad \rightarrow \quad \psi_1(R,t) = exp(-i\omega t) \cdot \psi_{1,0}(R)$ {$\omega$ - is the appropriate parameter of frequency}.

As for the case 1) above, equation (2.3) could be reduced to the proper *Riccati* type ODE of complex value (which has no solution in general case); it means that the general solution must be also of complex value [5]. Condition 2) leads to the real meanings of a solution (not of complex value), so it could be associated with the special case of eikonal solutions [8-9] to radial equation (2.3), the *Bessel* functions.

Thus, we have obtained the exact *non-stationary* solutions of Eq. (1.1):

$$\psi(R, \theta, \varphi, t) = \psi_1(R, t) \cdot \psi_2(\theta, \varphi),$$
$$U(R, \theta, \varphi, t) \sim U_1(R),$$

- where $\theta \in (0, \pi)$, the functions $\psi_2(\theta, \varphi)$, $\psi_1(R, t)$ are determined by Eqs. (2.2), (2.3) respectively. As for equation (2.3) of *Riccati* type, we should note that a modern method exists for obtaining of the numerical solution of *Riccati* equations with a good approximation [10-11].

Jumping of a phase-function of the component (2.2) for a wavefunction $\psi$ being equal to zero at the meaning of parameter $\theta = \pi/2$ or $\varphi = -C_2/C_1$, could be associated with the existence of an *optical vortex* [12] at this point. Optical vortex (also known as a screw dislocation or phase singularity) is a zero of an optical field, a point of zero intensity.
Research into the properties of vortices has thrived since a comprehensive paper [12], described the basic properties of "dislocations in wave trains".

## 3. Paraxial approximation.

As for the appropriate example of *paraxial* approximation for such an exact solution (2.2) of the full Schroedinger equation (1.1), it could be easily obtained in the case $\theta \to +0$:

Let us express the component of solution (2.2) in the Cartesian co-ordinates *x*, *y*, *z* as below:

$$R = \sqrt{x^2 + y^2 + z^2} = \sqrt{(r(x,y))^2 + z^2}, \quad \cos\theta = \frac{z}{\sqrt{(r(x,y))^2 + z^2}}, \quad \tan\varphi = \left(\frac{x}{y}\right),$$

$$\tan\left(\frac{\theta}{2}\right) = \frac{1-\cos\theta}{\sqrt{1-(\cos\theta)^2}} = \sqrt{\frac{1-\cos\theta}{1+\cos\theta}} = \sqrt{\frac{\sqrt{(r(x,y))^2 + z^2} - z}{\sqrt{(r(x,y))^2 + z^2} + z}},$$

$$\psi_2(\theta,\varphi) = (C_1\varphi + C_2)\cdot\ln\left(\tan\left(\frac{\theta}{2}\right)\right) = \tag{3.1}$$

$$= (C_1\cdot\arctan\left(\frac{x}{y}\right) + C_2)\cdot\ln\left(\frac{\sqrt{(r(x,y))^2 + z^2} - z}{\sqrt{(r(x,y))^2 + z^2} + z}\right) \cong \left(C_1\cdot\arctan\left(\frac{x}{y}\right) + C_2\right)\cdot\ln\left(\frac{\sqrt{x^2+y^2}}{2z}\right),$$

- where we assume $r(x, y) \ll z$ in (3.1) for the case of *paraxial* approximation of the component (2.2) of a solution; we could imagine such an approximation as below (Figs. 1-2), the meaning for variable z is chosen to be a large enough in regard to the meanings of variables $\{x, y\}$.

To avoid ambiguity, we should especially note that the partial case (2.2) as well as *paraxial* approximation for such a solution - present a very special (rare) case of real value solutions, which form a sub-class of exact solutions from the very wide class of complex value solutions of the *Riccati* equation (2.1). Besides, *paraxial* approximation (3.1) is meaningful for presenting a specific topology of a solution (2.2), see Figs.1-2 below.

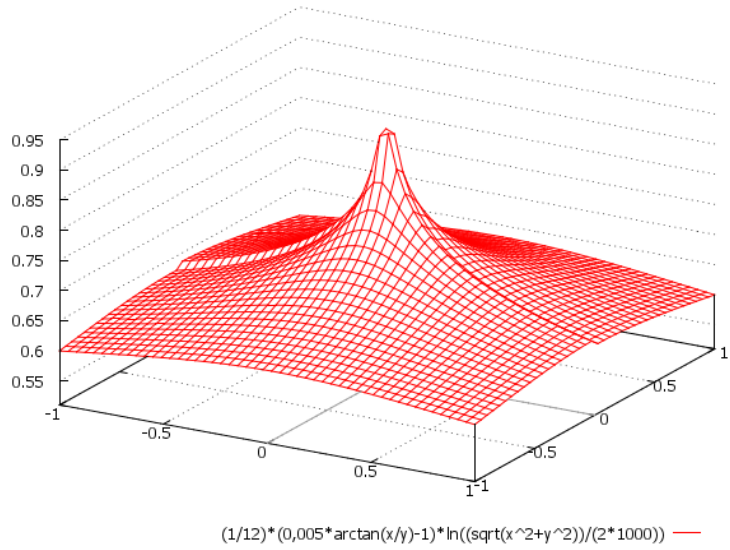

(1/12)*(0,005*arctan(x/y)-1)*ln((sqrt(x^2+y^2))/(2*1000))

Fig.1. A *schematic* plot of the component of a solution (2.2),

here we designate: $z = 1'000$, $\{x, y\} \in [-1, 1]$.

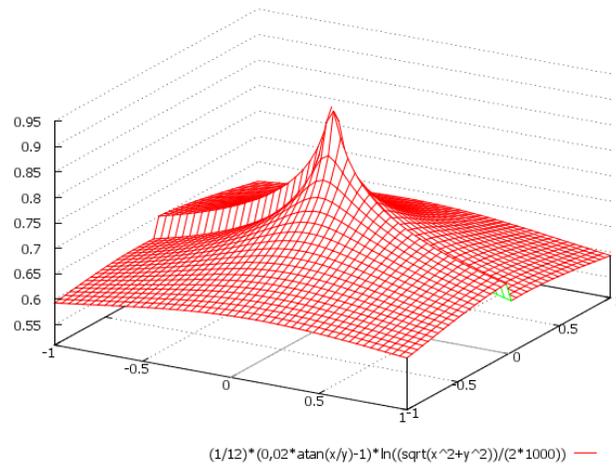

(1/12)*(0,02*atan(x/y)-1)*ln((sqrt(x^2+y^2))/(2*1000))

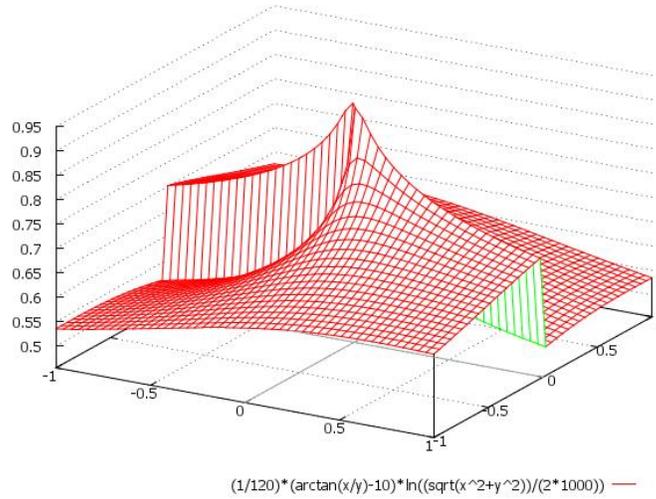

(1/120)*(arctan(x/y)-10)*ln((sqrt(x^2+y^2))/(2*1000))

Fig.2. A *schematic* plots of the component of a solution (2.2),

here we designate: $z = 1'000$, $\{x, y\} \in [-1, 1]$.

Let us also *schematically* imagine the component of a solution, associated with the part ~ $\ln(\tan(\theta/2))$ for the *paraxial* approximation $\theta \to +0$ at Fig.3 (to compare it with the solutions above).

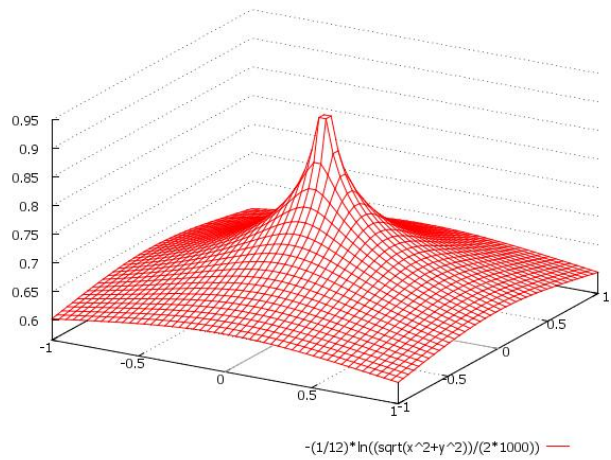

-(1/12)*ln((sqrt(x^2+y^2))/(2*1000))

Fig.3. A *schematic* plot of the component $\ln(\tan(\theta/2))$ of (2.2),

here we designate: $z = 1'000$, $\{x, y\} \in [-1, 1]$.

## 4. Discussion.

We should note that quantization conditions are not considered in this derivation; the main reason is that such a quantization transforms the radial part of solution to well-known Bessel-type integral invariants [5] with the discrete quantization as in harmonic oscillator case.

Nevertheless, the solutions presented are proved to be L2-integrable (which is a necessary condition for finite normalization and thus physical realization). Indeed, *Riccati*-type equations are known to have a solution which jumping [5] only at some finite (discrete) meanings of the range of variable argument (time-parameter, for example); such a jumping could be associated with the existence of a screw dislocation or phase singularity [12] at this point. For the (2.2)-type of solution, it means the existence of optical vortex at this point (point of zero intensity).

A very specific feature of Riccati-type equations as jumping of a solution at finite (discrete) meanings of the range of argument is especially outlined here just to orient the reader about the intent or purpose of the research (how the current development addresses any physical phenomenon – the optical vortex, for example, or "dislocations in wave trains" [12]).

Besides, we should take into consideration the requirement for the finite normalization and physical realization of a solution as below: - the square of wavefunction $\psi$ is assumed to be equal to the meaning of probability amplitude to detect of particle in a position space (per unit of volume). So, the meaning of wavefunction $\psi$ should be less than < 1 in any case.

Such a requirement is used for choosing of restrictions regarding the range of existence of a solution (i.e., appropriate normalization of a solution), see Figs.1-3.

Finally, we should spherically integrate the square of wavefunction $\psi$ over all the volume of position space, it must be equal to unit = 1 (this is basic postulate of quantum mechanics [3]); it let us choose the proper constants of a solution, according to the given initial conditions.

Also, the linear azimuthal dependence in equation (3.1) suggests that complete rotations (multiples of $2\pi$) of the sphere lead to different physical solutions; so, we should restrict the solution at the range of parameter $\varphi \in [0, 2\pi)$. Besides, inequality: $|C_1 \cdot \varphi + C_2| < 1$ should be valid for all meanings of function $(C_1 \cdot \varphi + C_2)$ in the chosen range of parameter $\varphi$ (see note above for meaning of probability amplitude), according to the given initial conditions.

Analogously, we should restrict the range of parameter $\theta \in (0, \pi)$ to the range $\theta \in [\theta_0, \theta_1]$ {where $\theta_0 = 2 \cdot \arctan(1/e) \cong 0{,}2244\pi$, $e = 2.71828...$, $\theta_1 = 2 \cdot \arctan(e) \cong 0{,}7756\pi$} for the reason that inequality: $|\ln \tan(\theta/2)| < |\tan(\theta/2)| \leq 1$ should be valid for all meanings of function $\ln(\tan(\theta/2))$ in the range of $\theta \in [\sim 40{,}4°, \sim 139{,}6°]$.

As for the importance or relevance of this development, a lot of authors have been executing their researches in quantum mechanics [3-4] to obtain the analytical solutions of Schroedinger equation. But there is an essential deficiency of *non-stationary* 3D solutions indeed; the elegant solution of such a type is proposed here for the full 3+1 dimensional space-time, non-relativistic Schroedinger equation in this derivation.

## 5. Conclusion.

The motivation of this derivation is to demonstrate the hidden possibilities of Schroedinger equation as partial solutions of *Riccati* type. Existence of such a solution means that wavefunction of particle could reveal the jumping at some moment; such a jumping could be associated with the existence of a screw dislocation or phase singularity at this point. It means the existence of optical vortex at this point (point of zero intensity).